\documentclass[preprint,eqsecnum,nofootinbib,natbib,aps]{revtex4}
\usepackage{graphicx}

\begin{document}

\title{Gravitational Wave Spectrum in Inflation with Nonclassical States}

\author{Seoktae Koh}
\email{kohst@ihanyang.ac.kr}
\author{Sang Pyo Kim}
\email{sangkim@kunsan.ac.kr}
\affiliation{Department of Physics, Kunsan National University, Kunsan 
573-701,Korea}
\author{Doo Jong Song}
\email{djsong@kao.re.kr}
\affiliation{Korea Astronomy Observatory, Daejeon 305-348, Korea}

\begin{abstract}
The initial quantum state during inflation may evolve to a highly
squeezed quantum state due to the amplification of the time-dependent
parameter, $\omega_{phys}(k/a)$, which may be the modified dispersion 
relation in trans-Planckian physics. This squeezed quantum state is a
nonclassical state that has no counterpart in the classical theory.
We have considered the nonclassical states such as squeezed,
squeezed coherent, and squeezed thermal states,
and calculated the power spectrum
of the gravitational wave perturbation when the mode leaves the horizon.
\end{abstract}

\maketitle

\section{Introduction}

Gravitational perturbations such as density perturbation and gravitational wave
would have been created from quantum fluctuations during the inflation period
when the wavelength of the perturbation was much smaller than
the horizon size \cite{hawking82}. 
And if the wavelength stretches out to the Hubble radius
during the inflation,
the amplitude would be frozen out. 
The power spectrum of the gravitational wave, 
if the initial state is an adiabatic vacuum 
state (Bunch-Davies vacuum) \cite{birrel}, 
gives a scale invariant one with $n_s =1$ and $n_T=0$ 
at the horizon crossing time, which is defined by \cite{liddle93}
\begin{eqnarray}
P_s = A_s k^{n_s-1}, \quad P_T = A_T k^{n_T},
\end{eqnarray}
where $A_s$ and $A_T$ are the amplitudes of scalar and tensor perturbations.

These gravitational wave and primordial density
perturbations may have undergone a squeezing process due to a time-varying
spacetime background \cite{grishchuk90}.  
If the Hamiltonian is given by a free Hamiltonian
term ($aa^{\dag}+a^{\dag}a$) and an interacting term ($aa + a^{\dag}a^{\dag}$)
in the Heisenberg picture 
where $a$ and $a^{\dag}$ are annihilation and creation operators, 
respectively, 
the squeezing occurs due to the interaction with a background.
Although the squeezing can result from the interaction with the 
background spacetime, it is also possible to squeeze the vacuum state through
a parametric interaction of the time-dependent
dispersion relation at the subhorizon scale. This
fact may be tested in the trans-Planckian physics.

The wavelength that corresponds
to the present horizon size, when dated back to the early universe,
 might be smaller than the Planck scale
\cite{martin01,brandenberger01}. 
Though the Planck scale physics has not been completely understood, 
several ways were suggested to treat the trans-Planckian effects
such as modifying dispersion relations, $\omega_{phys}(k/a)$ 
\cite{martin01,brandenberger01}, 
considering non-vacuum initial state \cite{hui01,goldstein03}, 
and modifying the space-time
commutation relation motivated from the string theory \cite{kempf01}.
In Ref. \cite{hui01}, a squeezed vacuum state as an initial
condition was used to calculate the consistency relations ($r \equiv A_T/A_s$)
which is the ratio of the power spectra
between scalar and tensor perturbations in inflation.

In this paper we shall consider in trans-Planckian physics 
whether the initial state may
evolve to a squeezed quantum state due to the parametric interaction
which comes from the modified dispersion relation. So 
the gravitational wave in the inflation period may exist, in general, 
in a squeezed 
quantum state, not the vacuum state. This squeezing process may 
slightly  modify the power spectrum of the gravitational wave 
perturbations for superhorizon scales by the order of $\frac{H^2}{M^2}$
where $M$ is the cutoff scale.
We shall consider several types of nonclasscial states 
such as squeezed, squeezed coherent, and
squeezed thermal states. The coherent states are in the borderline
between quantum and classical states. The expectation value of the 
fields with respect to coherent states  follow the classical
equations of motion. So the coherent states may be a convenient tool to 
deal with the classicality of the quantum fluctuations. Further,
it may be not excluded that the universe would have been very hot from 
the beginning. This means that thermal effects would 
play an important role during the history of the early universe. Thus,  
thermal states may have some relevance 
in dealing with the gravitational perturbations.

Nonclassical states that have no counterpart in classical theory
are widely studied in quantum optics. Nonclassical states of light exhibit
the features such as sub-Poissonian photon statistics \cite{short83} and
squeezing \cite{slusher85}.
Quasiprobability distribution functions \cite{cahill69}, which  
play a similar, but not exactly the same, role of a classical probability,
are used to quantitatively measure the nonclassical behavior of states.
A particularly useful distribution is the Glauber-Sudarshan 
$P$ distribution that is
defined in terms of the density operator
\cite{sudarshan63}
\begin{eqnarray}
\hat{\rho} = \int d^2 \alpha P(\alpha) |\alpha\rangle\langle\alpha|,
\label{glauber}
\end{eqnarray}
where $|\alpha\rangle$ is a coherent state and $\alpha$ is a complex variable.
If the $P$ distribution of a state is more singular than a delta function
and is not a positive definite, the state is classified as 
a nonclassical state \cite{hillery85}. 
The regularity condition for the quasiprobability distribution function
is used to test whether states are either classical
or nonclassical.

The organization of this paper is as follows.
In Sec. II, we introduce the squeezed quantum states and 
quantize the gravitational wave perturbation in the inflation period. 
The squeeze parameter is found from the mode solution in the
trans-Planckian theory. In Sec. III,  we calculate 
the two-point correlation function in various
nonclassical states and obtain
the power spectrum of the 
gravitational wave perturbation for the superhorizon scale.
We finally conclude and discuss in Sec. IV.

\section{Quantization of scalar fields in de Sitter Space with Squeezed
Quantum State} \label{squeeze}

The gravitational wave perturbation in
the inflation period will be quantized
using the two-mode squeezed state formalism.
The metric of the spacetime in conformal time is given by
\begin{eqnarray}
ds^2 = a^2 (\eta)[-d\eta^2 +(\gamma_{ij}+h_{ij})dx^i dx^j],
\end{eqnarray}
where $h_{ij}$ is a traceless, transverse perturbation and is assumed to satisfy
$|h_{ij}| \ll \gamma_{ij}$. In the momentum space, $h_{ij}$ is expanded as
\begin{eqnarray}
h_{ij}(x)=\frac{1}{a}\int \frac{d^3 k}{(2\pi)^{3/2}} \sum e_{ij}
\mu_k(\eta) e^{i{\bf k}\cdot{\bf x}},
\end{eqnarray}
where $e_{ij}$ is a polarization tensor and $\mu_k$ is 
a complex function. Due to
the reality condition of $h_{ij}$, one has $\mu_{k}^{\ast} = \mu_{-k}$.

The perturbed action for the gravitational wave is
\begin{eqnarray}
\delta S &=& \int d^4 x \frac{a^2}{2}[h_{ij}^{\prime}h^{ij \prime}
-F(\nabla,a)h_{ij}h^{ij}],   \\
         &=& \int d\eta d^3 {\bf k} \left[\mu_{k}^{\prime}\mu_{-k}^{\prime}
-\frac{a^{\prime}}{a}(\mu_{k}^{\prime}\mu_{-k}+\mu_{k}\mu_{-k}^{\prime})
+\left(\left(\frac{a^{\prime}}{a}\right)^2
-\omega^2(k)\right)\mu_{k}\mu_{-k}\right], \label{action}
\end{eqnarray}
where a prime denotes the derivative with respect to the conformal time $\eta$,
and $F(\nabla, a(\eta))$ is a function of spatial derivative, $\nabla$, and 
the conformal time, and becomes $\omega (k)$ after Fourier transforming.
The last term in the second line, a modified dispersion relation, 
takes into account the 
trans-Planckian effect. It represents  a time-dependent 
non-linear property when the size of the
perturbation mode is much smaller than the Planck scale, 
\begin{eqnarray}
\omega^2 (k) = a^2 \omega_{phys}^2 (k/a),
\end{eqnarray}
but recovers the linear 
relation, $\omega \simeq k$, when the mode scale is greater than the 
Planck scale as in the standard inflationary model.
From Eq. (\ref{action}), one obtains the conjugate momentum
\begin{eqnarray}
\pi_{k} = \frac{\partial \mathcal{L}_k}{\partial \mu_{-k}^{\prime}}
=\mu_{k}^{\prime}-\frac{a^{\prime}}{a} \mu_k,
\end{eqnarray}
and the Hamiltonian density
\begin{eqnarray}
\mathcal{H}_k &=& \pi_{k}\mu_{-k}^{\prime}+\pi_{-k}\mu_{k}^{\prime}
 -\mathcal{L}_k, \nonumber \\
&=& \pi_{k}\pi_{-k}+\omega^2(k) \mu_{k}\mu_{-k}
 +\frac{a^{\prime}}{a}(\pi_{k}\mu_{-k}+\pi_{-k}\mu_{k}). 
\label{hamiltonian}
\end{eqnarray}

The Hamiltonian density leads to the equation of motion
\begin{eqnarray}
\mu_k^{\prime\prime}+\left(\omega^2(k)-\frac{a^{\prime\prime}}{a}\right)
\mu_k =0. \label{eom}
\end{eqnarray}
To quantize the fields we introduce 
two linear time-dependent operators \cite{malkin70,kim99}
\begin{eqnarray}
& & \hat{a}_{k}(\eta) = -i\left(\varphi^{\ast\prime}_k(\eta)
-\frac{a^{\prime}}{a}
\varphi^{\ast}_k (\eta)\right) \hat{\mu}_{k}
+i \varphi_{k}^{\ast}(\eta) \hat{\pi}_{k}, 
\nonumber \\
& & \hat{a}_{-k}^{\dag}(\eta) = i\left(\varphi^{\prime}_k(\eta)
-\frac{a^{\prime}}{a}
\varphi_k(\eta)\right) \hat{\mu}_{k}-i \varphi_{k}(\eta)
 \hat{\pi}_{k}. \label{operator}
\end{eqnarray}
Here we treated $\hat{\mu}_k$ and $\hat{\pi}_k$ as Schr\"{o}dinger 
operators. These 
operators satisfy the quantum Liouville-von Neumann equation
\cite{malkin70,kim99,kim01,kim00}
\begin{eqnarray}
i\frac{\partial}{\partial \eta} \hat{a}_k (\eta) + [\hat{a}_k(\eta),
\hat{\mathcal{H}}_k(\eta)] =0,
\end{eqnarray}
and the same equation holds for $\hat{a}^{\dag}_k$. 
It should be remarked that the
eigenstates of these operators are exact solutions of the time-dependent 
Schr\"{o}dinger equation, when $\varphi_{k}$ is a complex solution 
to the equation of motion
\begin{eqnarray}
\varphi_{k}^{\prime\prime} + \left(\omega^2(k)
-\frac{a^{\prime\prime}}{a}\right)\varphi_{k} =0.
\label{mode_eq}
\end{eqnarray}
The commutation relation of the fields in the momentum space at equal times 
\begin{eqnarray}
[\hat{\mu}_k,\hat{\pi}_{k^{\prime}}] = i \delta(k+k^{\prime}),
\end{eqnarray}
leads to the usual commutation relation
\begin{eqnarray}
[\hat{a}_k,\hat{a}_{k^{\prime}}^{\dag}] = \delta(k-k^{\prime}),
\end{eqnarray}
when the Wronskian condition
\begin{eqnarray}
\varphi_k \varphi_{k}^{\ast\prime}
-\varphi_{k}^{\ast}\varphi_{k}^{\prime} = i \label{wronskian}
\end{eqnarray}
is imposed.
We may thus interpret the $\hat{a}_k$ and $\hat{a}^{\dag}_{-k}$ as a 
time-dependent annihilation and creation operator. These operators 
do not, however, diagonalize the Hamiltonian (\ref{hamiltonian}), but
provide the exact quantum states of the time-dependent 
Schr\"{o}dinger equation \cite{kim01}.

There are two ways of squeezing quantum states. First, we choose some 
preferred complex solution to Eqs. (\ref{mode_eq}) and (\ref{wronskian}),
which may be chosen based on, for instance, the minimum particle 
creation postulate \cite{parker69} or the minimum uncertainty 
\cite{kim99}. Then a general solution to Eqs. (\ref{mode_eq})
and (\ref{wronskian}) is the superposition
\begin{eqnarray}
\varphi_{ks}(\eta) = u_0 \varphi_k(\eta) + 
v_0 \varphi_k^{\ast}(\eta),  \label{mode_sol2}
\end{eqnarray}
with
\begin{eqnarray}
|u_0|^2 -|v_0|^2 = 1.
\end{eqnarray}
Here the subscript $s$ denotes a squeezed state.
We may write $u_0$ and $v_0$ as squeeze parameters
\begin{eqnarray}
u_0 = \cosh r, \quad v_0 = e^{-i\phi} \sinh r.
\end{eqnarray}
Second, a given state may be squeezed due to a parametric  interaction, here,
time-dependent spacetime background. The preferred solution $\varphi_k$
at $\eta$  can be written as the superposition  of $\varphi_k$ and 
$\varphi_k^{\ast}$ at a different conformal time $\eta_0$. Note that
$\varphi_k(\eta)$ and $\varpi_k(\eta) (\equiv \varphi^{\prime}(\eta)-
\frac{a^{\prime}}{a}\varphi(\eta))$ satisfy the Hamilton  equations.
Then the two functions
\begin{eqnarray}
u_k(\eta,\eta_0) &=& i[\varpi_k (\eta)\varphi_k^{\ast}(\eta_0)-
\varphi_k(\eta)\varpi_k^{\ast}(\eta_0)], \nonumber \\
v_k(\eta,\eta_0) &=& -i [\varpi_k(\eta) \varphi_k(\eta_0)
-\varphi_k(\eta) \varpi_k(\eta_0)], \label{dynasqu}
\end{eqnarray}
satisfy Eq. (\ref{mode_eq}) at $\eta$ and $\eta_0$, and have the initial
value $u_k(\eta_0,\eta_0)=1$ and $v_k(\eta_0,\eta_0) =0$. The inverse
transformation  of Eq. (\ref{dynasqu}) is found to be
\begin{eqnarray}
\varphi_{ks}(\eta) &=& u_k(\eta,\eta_0) \varphi_k (\eta_0) + v_k(\eta,\eta_0)
\varphi_k^{\ast}(\eta_0), \nonumber \\
\varpi_{ks}(\eta) &=& u_k(\eta,\eta_0) \varpi_k(\eta_0) +
v_k(\eta,\eta_0) \varpi_k^{\ast}(\eta_0). \label{mode_sol}
\end{eqnarray}
In fact, $\varphi_{ks} (\eta)$ and $\varpi_{ks} (\eta)$ are solutions to Hamilton
equations with the initial data $\varphi_k(\eta_0)$ and $\varpi_k(\eta_0)$.

In the both cases, substituting Eqs. (\ref{mode_sol2}) and (\ref{mode_sol})
into Eq. (\ref{operator}), 
we obtain the Bogoliubov transformation
\begin{eqnarray}
\hat{a}_{ks}(\eta) &=& u^{\ast} \hat{a}_{k}(\eta) - 
v^{\ast} \hat{a}_{-k}^{\dag}(\eta), \label{ann_operator}\\
\hat{a}_{-ks}^{\dag}(\eta) &=& u \hat{a}_{-k}^{\dag}(\eta)
-v \hat{a}_{k}(\eta). \label{cre_operator}
\end{eqnarray}
In terms of the squeeze operator \cite{stoler70}
\begin{eqnarray}
\hat{S}_k(z,\eta) = \exp\left[
z_k \hat{a}^{\dag}_k(\eta) \hat{a}_{-k}^{\dag}(\eta)
-z_k^{\ast}\hat{a}_k(\eta)\hat{a}_{-k}(\eta)\right],
\end{eqnarray}
where $z_k=r_k e^{i\phi_k}$,
the Bogoliubov transformation can also be written as
\begin{eqnarray}
\hat{a}_{ks}(\eta) &=& \hat{S}_k(z,\eta)\hat{a}_{k}(\eta)
\hat{S}_k^{\dag}(z,\eta), \\
\hat{a}^{\dag}_{-ks}(\eta) &=& \hat{S}_k(z,\eta)\hat{a}^{\dag}_{-k}(\eta)
\hat{S}_k^{\dag}(z,\eta).
\end{eqnarray}
The squeeze operator $\hat{S}(z,\eta)$ maps a number state into a squeezed
number state which is an eigenstate of the squeezed number operator,
$\hat{N}_{ks}(\eta) 
\equiv \hat{a}^{\dag}_{ks}(\eta)\hat{a}_{ks}(\eta)$,
\begin{eqnarray}
|n,\eta\rangle_s = \hat{S}(z,\eta) |n,\eta\rangle,
\end{eqnarray}
where 
\begin{eqnarray}
|n,\eta\rangle_s = |n,z;{\bf k},{\bf -k},\eta\rangle.
\end{eqnarray}

We now solve the mode equation Eq. (\ref{mode_eq}) in
each region during the inflation period.  
Consider an exponential inflation with the scale factor given by
\begin{eqnarray}
a(\eta) =-\frac{1}{H\eta}.
\end{eqnarray}
When momenta $k$ are greater than the cutoff scale, $k_c \equiv aM$,
where $M$ is the cutoff scale, the dispersion relation $\omega(k)$ shows 
the non-linear property. Several forms of $\omega_{phys}(k/a)$ in this
region are suggested in literature 
\cite{martin01,brandenberger01,hui01,goldstein03,kempf01}. 
To solve the equation in this
region, we assume that the adiabatic approximation holds, $(|\omega^{\prime}|
\ll \omega^2)$, and get the WKB-type solution
\begin{eqnarray}
\varphi_{k}(\eta) = \frac{1}{\sqrt{2\omega}} e^{-i\int \omega(\eta^{\prime})
d\eta^{\prime}}.
\label{sol_1}
\end{eqnarray}
If $\omega_{phys}(k/a)$ is given by \cite{martin01}
\begin{eqnarray}
\omega_{phys}(k/a) = M \tanh^{1/p}\left[\left(\frac{k}{aM}\right)^p\right],
\label{dispersion}
\end{eqnarray}
where $p$ is an arbitrary coefficient,
the condition of the adiabatic approximation requires that $\frac{H}{M} \ll 1$.
In fact, the 
small amplitude of cosmic microwave background fluctuations constrains
$\frac{H}{M} \leq 10^{-5}$ while our present Hubble scale crossed 
the horizon during inflation \cite{niemeyer01}.
The mode crosses the cutoff scale when $k=k_c$.
If the wavelength of the mode is larger than the cutoff length scale, then the
dispersion relation becomes linear in the physical wavenumber, 
$\omega(k) \simeq
k$, and leads to the exact solution 
\begin{eqnarray}
\varphi_k(\eta) = \left(1-\frac{i}{k\eta}\right) 
\frac{e^{-ik\eta}}{\sqrt{2k}}. \label{sol_2}
\end{eqnarray}
The variances of the field and its conjugate momentum are calculated
with the solutions (\ref{sol_1}) and (\ref{sol_2}) to see whether
the squeezing occurs during the expansion of the universe. In the
trans-Planckian era, the variances with respect to the vacuum state are
\begin{eqnarray}
\Delta \mu_k^2 &\equiv& \langle\mu_k \mu_{-k}\rangle =\frac{1}{2\omega}, \\
\Delta \pi_k^2 &\equiv& \langle \pi_k\pi_{-k}\rangle 
=\frac{\omega}{2}\left(1+\frac{1}{\omega^2\eta^2}\right),
\end{eqnarray}
and
\begin{eqnarray}
\Delta \mu_k^2 \Delta \pi_k^2 =\frac{1}{4}\left(1
+\frac{1}{\omega^2 \eta^2}\right).
\end{eqnarray}
Note that the uncertainty of the gravitational wave field $h_{ij}$ and 
its conjugate momentum are related as
\begin{eqnarray}
\Delta h_k^2 \equiv \frac{1}{a^2}\Delta \mu_k^2, \quad
\Delta \Pi_k^2 \equiv a^2 \Delta \pi_k^2.
\end{eqnarray}
If $\omega$ is given in the form of Eq. (\ref{dispersion}), we can
approximate $\omega \simeq k_c = aM$ for $k\gg k_c$. Then the 
uncertainty in the field, $\Delta \mu_k$, increases as the universe expands,
whereas $\Delta \pi_k$ decreases. However, these have the minimum 
uncertainty as long as $\frac{H}{M} \ll 1$. So we can see that the
squeezing occurs due to not only the expansion of the universe but
the time-dependent parameter $\omega$ in the trans-Planckian era.
On the other hand,
for $k < k_c$, the variances of the field and momentum with respect to
the squeezed vacuum state are
\begin{eqnarray}
\Delta \mu_k^2 &=& \frac{1}{2k}\left(1+\frac{1}{k^2\eta^2}\right), \\
\Delta \pi_k^2 &=& \frac{k}{2},
\end{eqnarray}
and
\begin{eqnarray}
\Delta \mu_k^2 \Delta \pi_k^2 = \frac{1}{4}\left(1+\frac{1}{k^2\eta^2}\right).
\end{eqnarray}
The modes whose wavelengths stay inside the horizon $(k\eta \gg 1)$
have the minimum uncertainty. For the superhorizon scales $(k\eta \ll 1)$,
however, both the variance of the field, $\Delta \mu_k$, and the
uncertainty $\Delta \mu_k \Delta \pi_k$ increase with the conformal time.
The scalar field in inflation can be described classically if the uncertainty 
$\Delta \mu_k \Delta \pi_k$ is much larger than the minimum uncertainty 
\cite{guth85}. So in this sense
the gravitational wave can be treated as a classical
when the wavelength of the mode is much larger than the horizon length.
The Bogoliubov coefficients $u$ and $v$ can be calculated from Eq. 
(\ref{dynasqu})
where $\eta_0$ denotes the conformal time in a region in which 
the trans-Planckian effect becomes important, 
and $\eta$ is the time after the mode leaves the
trans-Planckian cutoff scale. 
We choose the WKB solution (\ref{sol_1}) as the preferred complex solution 
$\varphi_k(\eta_0)$ in Eq. (\ref{mode_sol}), 
which implies squeezing may occur
through the trans-Planckian region.
With the solutions (\ref{sol_1}) and (\ref{sol_2}) in each region,
$u$ and $v$ take the form
\begin{eqnarray}
u (\eta,\eta_0) &=& e^{-ik\eta+i\int \omega d\eta} \sqrt{\frac{k}{4\omega}}
\nonumber \\
& & \times\left[1-\frac{1}{k^2\eta^2}+\frac{\omega}{k} -\frac{i}{k\eta}
\left(1+\frac{\omega}{k}\right) 
+i\left(1-\frac{i}{k\eta}\right)\left(\frac{1}{k\eta}-\frac{1}{k\eta_0}\right)
\right], \\
v (\eta,\eta_0) &=& e^{-ik\eta-i\int \omega d\eta}\sqrt{\frac{k}{4\omega}}
\nonumber \\
& & \times\left[-1+\frac{1}{k^2\eta^2}+\frac{\omega}{k}+\frac{i}{k\eta}
\left(1-\frac{\omega}{k}\right)
-i\left(1-\frac{i}{k\eta}\right)\left(\frac{1}{k\eta}-\frac{1}{k\eta_0}\right)
\right].
\end{eqnarray}

\section{Power Spectrum with Nonclassical States} \label{power}

In this section we calculate the power spectrum of 
gravitational wave perturbation when the perturbation mode stretches
out to the Hubble radius during the inflation. 
The two-point correlation function 
at equal times
\begin{eqnarray}
\langle \Psi| f(\eta,{\bf x})f(\eta,{\bf x}+{\bf r})|\Psi \rangle 
\equiv \int^{\infty}_0 \frac{dk}{k}\frac{\sin kr}{kr}P_f (k), \label{twopoint}
\end{eqnarray}
gives the power spectrum defined by
\begin{eqnarray}
P_f(k) &=& \frac{k^3}{2\pi^2}\int d^3 r e^{-i{\bf k}\cdot{\bf r}}
\langle \Psi|f({\bf x})f({\bf x}+{\bf r})|\Psi \rangle \\
&=&\frac{k^3}{2\pi^2}\langle |f_k|^2 \rangle .
\end{eqnarray}
Further, we consider the nonclassical states such as squeezed,
squeezed coherent, and squeezed thermal states instead of the vacuum.
 Nonclassical states are quantum states that have no classical analog. 
These states have the 
quasiprobability distributions which may not always compatible 
with the interpretation as 
a probability distribution \cite{cahill69}. The quasiprobability 
distribution $W(\alpha,s)$ is used to show the nonclassical behavior
quantitatively and the existence of negative regions of distribution 
for nonclassical states.
 The quasiprobability distribution, 
for example, the Glauber-Sudarshan $P(\alpha)$, is obtained 
from the complex Fourier transformation of Eq. (\ref{glauber})
\cite{cahill69}. In Appendix \ref{quasiprobability} we briefly review
the properties of the quasiprobability distribution, $W(\alpha,s)$.

\subsection{squeezed state}

First we consider the squeezed quantum state. In the previous section, we 
described the properties of the squeezed state. 
The ground squeezed state has a Gaussian wave function
\begin{eqnarray}
\Psi^{(s)}_0(\mu_k,\mu_{-k},\eta)= \mathcal{N} \exp 
\left[-i\frac{\varpi^{\ast}_{ks}}{\varphi^{\ast}_{ks}} |\mu_k|^2\right],
\end{eqnarray}
which is derived from the fact that $a_{ks}|0,\eta\rangle_s =0$. Here
$\mathcal{N}$ is a normalization factor.
On the other hand, excited squeezed states (squeezed number state)
have the non-Gaussian wave functions \cite{lesgourgues97}
\begin{eqnarray}
\Psi_n^{(s)} (\mu_k,\mu_{-k},\eta)=(-1)^n 
\left(\frac{\varphi_k}{\varphi_k^{\ast}}
\right)^n {Ln}\left(\frac{|\mu_k|^2}{|\varphi_k|^2}\right)\Psi_0^{(s)},
\end{eqnarray}
where $Ln$ is a Laguerre polynomial.
The temperature correlation functions of excited squeezed states show the 
nongaussianity.

To calculate the power spectrum with the squeezed quantum state, we use
the following relations
\begin{eqnarray}
_s\langle \hat{a}_k \hat{a}_{k^{\prime}} \rangle_s &=& u_k v_k^{\ast}
(1+2 n_k)\delta(k+k^{\prime}), \\
_s\langle \hat{a}_{-k}^{\dag}\hat{a}_{-k^{\prime}}^{\dag} \rangle_s 
&=& -u_k^{\ast} v_k (1+2n_k) \delta(k+k^{\prime}), \\
_s\langle \hat{a}_k \hat{a}_{-k^{\prime}}^{\dag}\rangle_s &=& |u_k|^2 (1+n_k)
\delta(k+k^{\prime}) +|v_k|^2 n_k\delta(k+k^{\prime}), \\
_s\langle \hat{a}_{-k}^{\dag} \hat{a}_{k^{\prime}} \rangle_s &=& 
|u_k|^2 n_k \delta(k+k^{\prime})
+|v_k|^2 (1+n_k)\delta(k+k^{\prime}),
\end{eqnarray}
where the inverse transformation of Eq. (\ref{operator}) was used
\begin{eqnarray}
& & \hat{a}_k = u  \hat{a}_{ks} + v^{\ast} \hat{a}_{-ks}^{\dag}, \nonumber \\
& & \hat{a}_{-k}^{\dag} = u^{\ast} \hat{a}_{-ks}^{\dag} + v \hat{a}_{ks}.
\end{eqnarray}
Then the two point correlation function of $h_{ij}$ with respect to the
squeezed quantum state is 
\begin{eqnarray}
_s\langle n,\eta| h_{ij}({\bf x}) h^{ij}({\bf x}+{\bf r})| n,\eta\rangle_s &=&
\frac{1}{a^2}\int \frac{d^3 k}{(2\pi)^{3/2}} 
\frac{d^3 k^{\prime}}{(2\pi)^{3/2}} 
 e^{i{\bf k}\cdot{\bf x}}e^{i{\bf k^{\prime}}\cdot 
({\bf x}+{\bf r})} \nonumber \\
& &\times
 _s\langle n,\eta | \mu_k(\eta)\mu_{k^{\prime}}(\eta)
|n,\eta \rangle_s \\
&=& \frac{1}{a^2}\int \frac{d^3 k}{(2\pi)^3} 
 e^{i{\bf k}\cdot{\bf r}} |\varphi_k|^2(1+2 n_k) 
\nonumber \\
& &\times [|u_k|^2+|v_k|^2+2 {\mathcal Re}(u_k^{\ast}v_k)].
\end{eqnarray} 
From this two-point correlation function, we can read off the power
spectrum of the gravitational wave 
\begin{eqnarray}
P_{gw}(k) =\frac{k^3}{2\pi^2}\frac{|\varphi_k|^2}{a^2}
(1+2 n_k)[(|u_k|^2+|v_k|^2) + 2 {\mathcal Re}(u_k^{\ast}v_k)].
\label{power}
\end{eqnarray}
This gravitational wave power spectrum may be rewritten,  using the 
squeezed mode $\varphi_{ks}$ in Eq. (\ref{mode_sol}), as
\begin{eqnarray}
P_{gw}(k) = \frac{k^3}{2\pi^2}\frac{|\varphi_{ks}|^2}{a^2}(1+2 n_k).
\end{eqnarray}
For $\eta \geq \eta_c$, $\omega \simeq k$ and $k\eta_c =\frac{M}{H}$,
one has 
\begin{eqnarray}
|u|^2 &=& \cosh^2 r \simeq 1 + \frac{H^4}{4 M^4}, \nonumber \\
|v|^2 &=& \sinh^2 r \simeq \frac{H^4}{4 M^4},  \nonumber \\
{\mathcal Re}(u^{\ast}v) &=& \frac{1}{2} \sinh 2r \cos \phi
 \simeq \frac{H^2}{2 M^2}\cos \left(
\frac{2M}{H}\right),  \label{bogoliubov}
\end{eqnarray}
where we have used the mode crossing the cutoff scale at $\eta=\eta_c$.
The growing mode solution of Eq. (\ref{mode_eq}) for $k \ll aH$ 
is given by 
$\varphi_k(\eta) \propto a(\eta)$.
Thus, the resulting power spectrum for the superhorizon sized perturbations
is
\begin{eqnarray}
P_{gw}(k) = \left(\frac{H}{2\pi}\right)^2 (1+2 n_k)
\left[1+\frac{H^2}{M^2}\cos \left(\frac{2M}{H}\right)\right].
\label{second order}
\end{eqnarray}
The power spectrum is corrected by the order $\frac{H^2}{M^2}$ and
this result differs from Ref. \cite{danielsson02}
but is similar to Ref. \cite{kaloper02} in order of magnitude. 
This correction is too
small to be detected by the CMB observations, but linear
order corrections in $(\frac{H}{M})$ are expected to 
be detectable \cite{bergstrom02}.
We use the result of Eq. (\ref{condi}) to show that the squeezed vacuum state
($n_k=0$) with the squeeze parameter 
in Eq. (\ref{bogoliubov}) is a nonclassical state 
with the critical value (see Appendix A)
\begin{eqnarray}
s_c = e^{-2 r} \simeq 1- \frac{H^2}{M^2}.
\end{eqnarray}
This satisfies the
condition for a state to be nonclassical $ (0\leq s_c \leq 1$), which
implies that the Glauber-Sudarshan
$P$ function ($s=1$) is more singular than a delta function.

We may consider a more general initial state in the trans-Planckian period
given by a squeezed state which is obtained from Eq. (\ref{mode_sol2})
\begin{eqnarray}
\varphi_{ks}(\eta_0) = u_0 \varphi_k (\eta_0) + v_0 \varphi_k^{\ast}
(\eta_0).
\end{eqnarray}
Here $\varphi_k(\eta_0)$ is the solution in Eq. (\ref{sol_1}).
Then the power spectrum for the superhorizon scales is
\begin{eqnarray}
P_{gw}(k) &\simeq& \left(\frac{H}{2\pi}\right)^2 (1+2 n_k)
\Biggl[|u_0|^2 + |v_0|^2 +\frac{H^2}{M^2} \Biggl(|u_0|^2
\cos\left(\frac{2M}{H}\right)  \nonumber \\
& & +|v_0|^2\cos \left(\frac{2M}{H}-\phi_k\right)
-2|u_0||v_0| \cos \left(\frac{2M}{H}-2\phi\right)\Biggr) \Biggr].
\end{eqnarray}
The terms containing the squeeze phase $\phi_k$ may be set to zero due to
the random character of the phase $\phi_k$ of each mode \cite{prokopec93}
to yield
\begin{eqnarray}
P_{gw}(k) \simeq \left(\frac{H}{2\pi}\right)^2 (1+2 n_k)
\left[|u_0|^2 + |v_0|^2 +|u_0|^2 \frac{H^2}{M^2}
\cos\left(\frac{2M}{H}\right)\right].
\end{eqnarray}
There is an increase of the overall factor by $|u_0|^2 + |v_0|^2$,
which may be absorbed into the normalization of the spectrum. 
Once this normalization is done, the coefficient of the correction term
can not be greater than one, thus slightly changing Eq. 
(\ref{second order}). Note that the maximally 
squeezed states, $r_k \rightarrow \infty$, 
of the adiabatic vacuum have a factor $1/2$.

\subsection{squeezed coherent state}

Now we consider the  squeezed coherent state. 
The squeezed coherent state is defined as an eigenstate of the 
annihilation operators $\hat{a}_{ks}(\eta)$ and $\hat{a}_{-ks}(\eta)$
\begin{eqnarray}
& & \hat{a}_{ks}(\eta)|\alpha,\eta\rangle_s 
= \alpha_k|\alpha,\eta\rangle_s,  \\
& & \hat{a}_{-ks}(\eta) |\alpha, \eta\rangle_s
= \alpha_{-k}|\alpha,\eta \rangle_s,
\end{eqnarray}
where $\alpha_k$ and $\alpha_{-k}$ are complex constants, and 
\begin{eqnarray}
|\alpha,\eta\rangle_s = |\alpha_k,\alpha_{-k},z,\eta\rangle.
\end{eqnarray}
The squeezed coherent state can be treated
algebraically by introducing a displacement operator
\begin{eqnarray}
\hat{D}(\alpha) &\equiv& \hat{D}(\alpha_k)\hat{D}(\alpha_{-k}), 
\nonumber \\
&=&  e^{\alpha_k\hat{a}_{ks}^{\dag}(\eta)
-\alpha_k^{\ast}\hat{a}_{ks}(\eta)}
e^{\alpha_{-k}\hat{a}_{-ks}^{\dag}(\eta)-\alpha_{-k}^{\ast}
\hat{a}_{-ks}(\eta)},
\end{eqnarray}
which is unitary
\begin{eqnarray}
\hat{D}(\alpha)\hat{D}^{\dag}(\alpha) =\hat{D}^{\dag}(\alpha)\hat{D}(\alpha)
=\hat{I}.
\end{eqnarray}
The displacement operator translates by constant amounts the annihilation 
and creation operators through the unitary transformation
\begin{eqnarray}
& &\hat{D}^{\dag}(\alpha)\hat{a}_{ks}(\eta)\hat{D}(\alpha) 
= \hat{a}_{ks}(\eta)+\alpha_k,
\nonumber \\
& &\hat{D}^{\dag}(\alpha)\hat{a}_{-ks}(\eta)\hat{D}(\alpha)
= \hat{a}_{-ks}(\eta)+\alpha_{-k}. 
\end{eqnarray}
Similar relations hold for $\hat{a}_{ks}^{\dag}$ 
and $\hat{a}_{-ks}^{\dag}$.
Hence one can see that the  squeezed coherent state results from the unitary 
transformation of the squeezed state
\begin{eqnarray}
|\alpha,\eta\rangle_s = \hat{D}(\alpha)|\eta\rangle_s = \hat{D}(\alpha)
\hat{S}(z,\eta)|\eta \rangle.
\end{eqnarray}
The expectation values of $\mu_k$ and $\pi_k$ with respect to the 
squeezed coherent state are 
\begin{eqnarray}
\mu_{kc} &\equiv& _s\langle \alpha, \eta|\hat{\mu}_k
|\alpha,\eta\rangle_s 
=~_s\langle \alpha,\eta| (\hat{a}_k\varphi_k +\hat{a}_{-k}^{\dag} 
\varphi_{-k}^{\ast})|\alpha,\eta \rangle_s \nonumber \\
&=& (u_k \alpha_k+v_{-k}^{\ast}\alpha_{-k}^{\ast})\varphi_k
+(u_{-k}^{\ast}\alpha_{-k}^{\ast}+v_k \alpha_k)\varphi_{-k}^{\ast}, \\
\pi_{kc} &\equiv& _s \langle \alpha, \eta|\hat{\mu}_k
|\alpha,\eta\rangle_s
=~_s\langle \alpha, \eta |\left(\hat{\mu}_{k}^{\prime}-\frac{a^{\prime}}{a}
\hat{\mu}_k \right)|\alpha, \eta \rangle_s \nonumber \\
&=& \mu_{kc}^{\prime}-\frac{a^{\prime}}{a}\mu_{kc}.
\end{eqnarray}
Note that $\mu_{kc}(\eta)$ is the solution of Eq. (\ref{mode_eq}).

We calculate the following relations with respect to the squeezed
coherent state 
\begin{eqnarray}
_s \langle \alpha,\eta | \hat{a}_k \hat{a}_{k^{\prime}}
|\alpha, \eta\rangle_s &=& [u_k^2 \alpha_k^2 +{v_{-k}^{\ast}}^2
{\alpha_{-k}^{\ast}}^2]\delta(k-k^{\prime}) +u_kv_k^{\ast}
\delta(k+k^{\prime}) \nonumber \\
& & +2 u_k v_{-k}^{\ast}
\alpha_{-k}^{\ast}\alpha_k \delta(k-k^{\prime}),  \\
_s \langle \alpha, \eta | \hat{a}_{-k}^{\dag} \hat{a}_{-k^{\prime}}^{\dag}
|\alpha, \eta\rangle_s &=& [{u_{-k}^{\ast}}^2 
{\alpha_{-k}^{\ast}}^2 +v_k^2
\alpha_k^2]\delta (k-k^{\prime}) +u_k^{\ast} v_k 
\delta(k+k^{\prime}) \nonumber \\
& &+2 u_{-k}^{\ast} v_k\alpha_{-k}^{\ast} \alpha_k \delta(k-k^{\prime}), \\
_s \langle \alpha, \eta | \hat{a}_k \hat{a}_{-k^{\prime}}^{\dag}|\alpha,\eta
\rangle_s &=& |u_k|^2 \delta(k+k^{\prime}) 
+(u_k u_{-k}^{\ast}+v_k v_{-k}^{\ast}) \alpha_k \alpha_{-k}^{\ast}
\delta (k-k^{\prime}) \nonumber \\
& &+[u_k v_k \alpha_k^2 
+u_{-k}^{\ast} v_{-k}^{\ast} {\alpha_{-k}^{\ast}}^2] 
\delta (k-k^{\prime}), \\
_s \langle \alpha, \eta | \hat{a}_{-k}^{\dag} 
\hat{a}_{k^{\prime}}|\alpha, \eta
\rangle_s &=& |v_k|^2 \delta(k+k^{\prime})+
(u_{-k}^{\ast}u_k +v_{-k}^{\ast} v_k)\alpha_{-k}^{\ast}\alpha_k
\delta (k-k^{\prime}) \nonumber \\
& &+(u_k v_k \alpha_k^2 +u_{-k}^{\ast} v_{-k}^{\ast}{\alpha_{-k}}^2
)\delta (k-k^{\prime}).
\end{eqnarray}
With these relations, we find the two point correlation function 
of $h_{ij}$ with respect to the squeezed coherent state
\begin{eqnarray}
_s \langle \alpha,\eta|h_{ij}({\bf x})h^{ij}({\bf x}+{\bf r})|
\alpha,\eta \rangle_s &=& \frac{1}{a^2}\int \frac{d^3 k}{(2\pi)^3}
e^{i{\bf k}\cdot{\bf r}}  \nonumber \\
&\times&  [|\varphi_k|^2 \{ (|u_k|^2+|v_k|^2)
+2 {\mathcal Re}(u_k^{\ast} v_k)\}
+e^{2 i{\bf k}\cdot {\bf x}} \mu_{kc}^2].
\end{eqnarray}
We finally obtain the power spectrum 
\begin{eqnarray}
P_{gw} = \frac{k^3}{2\pi^2 a^2}[|\varphi_k|^2\{(|u_k|^2
+|v_k|^2) +2{\mathcal Re}(u_k^{\ast} v_k)\} 
+e^{2i{\bf k}\cdot {\bf x}}\mu_{kc}^2].
\end{eqnarray}
The power spectrum consists of two parts; one is the
contribution from the squeezed vacuum state as in Eq. (\ref{power}) for
$n_k=0$
 and the other is a classical part
which is the expectation value with respect to the  squeezed coherent state.
The classical part $\mu_{kc}$ satisfies the classical equation of motion
 (\ref{eom}). As shown in Eq. (\ref{condi}), the regularity 
condition does not depend on the amplitude, $\alpha$,
of the squeezed coherent states so these states
also satisfy this condition as in the case of squeezed vacuum states.

\subsection{squeezed thermal state}

We now use thermofield dynamics (TFD) for time-dependent system 
\cite{takahasi75,spkim03}
and calculate the power spectrum of gravitational 
wave in a squeezed thermal state.
The thermofield dynamics is a canonical formalism for finite temperature
theory to describe quantum systems in thermal equilibrium. The idea of TFD
is to double the system by adding a fictitious system and extend the thermal
equilibrium in the system's Hilbert space to a thermal state, a pure state,
in the extended Hilbert space of the total system. 
We briefly introduce the TFD formalism in one-mode state and after that
calculate the power spectrum using the two-mode state TFD formalism.
The density operators
are defined by
\begin{eqnarray}
\rho_k(\eta) &=& \frac{1}{Z} 
e^{-\beta \omega_k \hat{a}_k^{\dag}(\eta)\hat{a}_k(\eta)}, \nonumber \\
\tilde{\rho}_k(\eta) &=& \frac{1}{Z} 
e^{-\beta \omega_k \tilde{\hat{a}}_k^{\dag}(\eta)\tilde{\hat{a}}_k(\eta)},
\label{density}
\end{eqnarray}
where 
\begin{eqnarray}
Z ={\rm Tr}e^{-\beta \mathcal{H}_k}.
\end{eqnarray}
Here $\beta$ and $\omega$ are constants that may be fixed by the initial
temperature and frequency. The density operator in the extended Hilbert space
is given by
\begin{eqnarray}
\hat{\rho}_k(\eta)=\rho_k(\eta) \otimes \tilde{\rho}_k(\eta) =
\frac{1}{Z^2}e^{-\beta\omega_k (\hat{a}_k^{\dag}(\eta)\hat{a}_k(\eta)
-\tilde{\hat{a}}_k^{\dag}(\eta)\tilde{\hat{a}}_k(\eta))}.
\end{eqnarray}
The thermal expectation value of the operator $\hat{A}$ of the system
now takes the form
\begin{eqnarray}
\langle A \rangle = {\rm Tr} \rho_k(\eta) \hat{A} = \langle 0(\beta),\eta|
\hat{A} |0(\beta),\eta\rangle,
\end{eqnarray}
where the thermal vacuum state is given by
\begin{eqnarray}
|0(\beta),\eta\rangle =\sqrt{1-e^{-\beta \omega_k}}e^{-(\beta \omega_k/2)
\hat{a}_k^{\dag}(\eta)\tilde{\hat{a}}_k^{\dag}(\eta)}|0,\eta\rangle ,
\end{eqnarray}
with $|0,\eta\rangle = |0,\tilde{0},\eta\rangle$.
As $\{ \hat{a}_k,\hat{a}_k^{\dag}\}$ and $\{ \tilde{\hat{a}}_k,
\tilde{\hat{a}}_k^{\dag}\}$ describe two independent systems, the nonzero
commutation relations become
\begin{eqnarray}
[\hat{a}_k,\hat{a}_{k^{\prime}}^{\dag}] =\delta(k-k^{\prime}), \quad
[\tilde{\hat{a}}_k,\tilde{\hat{a}}_{k^{\prime}}^{\dag}] =
\delta (k-k^{\prime}).
\end{eqnarray}
The thermal state is written as a time-dependent squeezed state
of the vacuum state
\begin{eqnarray}
|0(\beta),\eta\rangle = \hat{T}(\theta)|0,\eta\rangle,
\end{eqnarray}
where
\begin{eqnarray}
\hat{T}(\theta) = \exp[- \theta_k(\beta)\{\tilde{\hat{a}}_k(\eta)
\hat{a}_k(\eta)
-\hat{a}_k^{\dag}(\eta)\tilde{\hat{a}}_k^{\dag}(\eta)\}].
\end{eqnarray}
Here $\theta_k(\beta)$ is a temperature-dependent parameter determined by
\begin{eqnarray}
\cosh \theta_k(\beta) &=& (1-e^{-\beta \omega_k})^{-1/2}, \nonumber \\
\sinh \theta_k(\beta) &=& e^{-\beta \omega_k/2}(1-e^{-\beta \omega_k})^{-1/2}.
\label{theta}
\end{eqnarray}

In the two-mode state formalism the thermal operator $\hat{T}(\theta)$
can be written as
\begin{eqnarray}
\hat{T}(\theta) = \exp[- \theta_k(\beta)
\{\hat{a}_{ks} \tilde{\hat{a}}_{ks}
-\hat{a}_{ks}^{\dag}\tilde{\hat{a}}_{ks}^{\dag}\}]
\exp[-\theta_k(\beta)\{\hat{a}_{-ks}\tilde{\hat{a}}_{-ks}
-\hat{a}_{-ks}^{\dag}\tilde{\hat{a}}_{-ks}^{\dag}\}].
\end{eqnarray}
Then the time- and temperature-dependent annihilation and creation 
operators through the Bogoliubov transformation become
\begin{eqnarray}
\hat{a}_k(\beta, \eta) &=& \hat{T}(\theta) a_{ks} \hat{T}^{\dag}(\theta) 
= \cosh \theta_k(\beta) \hat{a}_{ks}(\eta)
-\sinh \theta_k(\beta)\tilde{\hat{a}}_{-ks}^{\dag}(\eta), \\
\tilde{\hat{a}}_k (\beta, \eta) &=& \hat{T}(\theta) \tilde{\hat{a}}_{ks}
\hat{T}^{\dag}(\theta) = 
\cosh \theta_k(\beta) \tilde{\hat{a}}_{ks} (\eta)
-\sinh \theta_k(\beta) \hat{a}_{-ks}^{\dag}(\eta).
\end{eqnarray}
We get similar equations for $\hat{a}_k^{\dag}(\beta,\eta),
\tilde{\hat{a}}_k^{\dag}(\beta,\eta)$ by using the Hermitian conjugate
of these equations. The thermal state is a time- and temperature-dependent
vacuum
\begin{eqnarray}
\hat{a}_{k}(\beta,\eta) |0(\beta),\eta\rangle_s 
= \tilde{\hat{a}}_{k}(\beta,\eta)
|0(\beta),\eta\rangle_s = 0.
\end{eqnarray}
The thermal state $|0(\beta),\eta\rangle_s$, as an eigenstate of the invariant
operators $\hat{a}_k(\beta,\eta)$, is an exact eigenstate of the total system.
To fix $\omega_k$ in Eq. (\ref{density}), we assume that at $\eta=\eta_i$
the universe was in thermal equilibrium state.
And if we neglect the expansion of the spacetime
at $\eta=\eta_i$, the  Hamiltonian is approximated as 
\begin{eqnarray}
\mathcal{H}_k=\pi_k \pi_{-k} + \omega_k \mu_k \mu_{-k},
\end{eqnarray}
to which we can set $\omega_k =
a(\eta_i) \omega_{phys}(k/a(\eta_i))$.

Using a similar method as in the previous sections, 
we get the following relations
\begin{eqnarray}
_s \langle \theta(\beta),\eta |\hat{a}_k 
\hat{a}_{k^{\prime}}| \theta(\beta),\eta
\rangle_s &=& u_k v_k^{\ast}(\cosh^2 \theta_k+\sinh^2 \theta_k)
 \delta(k+k^{\prime}), \\
_s \langle \theta(\beta),\eta | \hat{a}_{-k}^{\dag} 
\hat{a}_{-k^{\prime}}^{\dag}|
\theta(\beta),\eta\rangle_s &=&
u_k^{\ast} v_k (\cosh^2 \theta_k +\sinh^2 \theta_k)\delta(k+k^{\prime}), \\
_s \langle \theta(\beta),\eta | 
\hat{a}_k \hat{a}_{-k^{\prime}}^{\dag}
| \theta(\beta),\eta \rangle_s &=&
(|u_k|^2 \cosh^2 \theta_k + |v_k|^2 \sinh^2 \theta_k) \delta(k+k^{\prime}), \\
_s \langle \theta(\beta),\eta | \hat{a}_{-k}^{\dag} 
\hat{a}_{k^{\prime}}| \theta(\beta),\eta \rangle_s &=&
(|u_k|^2 \sinh^2 \theta_k + |v_k|^2 \cosh^2 \theta_k) \delta(k+k^{\prime}).
\end{eqnarray}
Using the above relations,  we can find the two-point correlation 
functions with the squeezed thermal state 
\begin{eqnarray}
_s\langle \theta(\beta),\eta| h_{ij}({\bf x})h^{ij}({\bf x}+{\bf r})|
\theta(\beta),\eta \rangle_s &=&
\frac{1}{a^2}\int \frac{d^3 k}{(2\pi)^3}
e^{i{\bf k}\cdot {\bf r}}|\varphi_k|^2  \nonumber \\
&\times& [(|u_k|^2+|v_k|^2) 
+ 2 {\mathcal Re}(u_k v_k)](1+ 2 \sinh^2 \theta_k),
\end{eqnarray}
from which we finally obtain the power spectrum
\begin{eqnarray}
P_{gw}(k) = \frac{k^3}{2\pi^2} \frac{|\varphi_k|^2}{a^2}
 [(|u_k|^2+|v_k|^2)
+ 2{\mathcal Re}(u_k v_k)](1+2 \sinh^2 \theta_k).
\end{eqnarray}
Here $\theta_k$ follows from Eq. (\ref{theta}) as determined by 
the Bose-Einstein distribution at the initial time:
\begin{eqnarray}
\sinh^2 \theta_k = \frac{1}{e^{\beta \omega_k}-1}.
\end{eqnarray}
Thus the resulting power spectrum with the squeezed thermal state  is 
corrected by the thermal contribution factor  $(1+2\sinh^2 \theta_k)$.
The condition of the quasiprobability distribution $W(\alpha,s)$ being
regular for the squeezed thermal state is from (\ref{condi}), 
\begin{eqnarray}
s < s_c = \frac{1}{1-e^{-\beta \omega_k}}\left(1-\frac{H^2}{M^2}\right),
\end{eqnarray}
where we have used the relations (\ref{bogoliubov}). In the limit of 
zero temperature, $\beta \rightarrow
\infty$, this condition reduces to that of the squeezed
vacuum state. However, in the high temperature limit,
$\beta \rightarrow 0$,
this state may be regarded as a classical state.

\section{Conclusion and Discussion} \label{summary}

We have studied the spectrum of gravitational wave perturbations 
in nonclassical states during inflation.
Gravitational wave or primordial density perturbation, which is responsible
for the large scale structure formation, would have been
generated from quantum
fluctuations when the wavelengths of the perturbation modes were much smaller
than the horizon scales. After the wavelengths were stretched out to the
horizon scale during inflation, they could be treated as a 
classical perturbation.

Any state can be, in the cosmological expansion, 
highly squeezed
and this phenomenon is quite a quantum effect which cannot be explained in 
classical theory. Squeezing can occur when the
time-dependent parameter is amplified at the subhorizon scale. We considered
in this paper that an initial state can evolve to its highly squeezed
state due to the amplifying time-dependent parameter which plays a role 
of the modified dispersion relation in trans-Planckian physics.
So the gravitational field may exist,in general,
in a squeezed quantum state during inflation. 
This is a dynamical process occurring through Eq. (\ref{mode_sol}).
In Eq. (\ref{mode_sol}), the Bogoliubov coefficients $u$ and $v$ are functions
of $\eta$ and $\eta_0$ and satisfy the Eq. (\ref{mode_eq}). From this
point of view, the underlying physical concept differs from 
that of Ref. \cite{hui01,goldstein03}. Actually
in Ref. \cite{hui01,goldstein03}, 
since the squeezed state was put as an initial 
condition, there was not a dynamical squeezing process. 
On the contrary, in this
paper any initial state is squeezed through the dynamical process due to
the time-dependent parameter.

We also considered the nonclassical state whose behavior can be quantitatively 
described by the quasiprobability distribution such as
squeezed quantum state, squeezed coherent state, and squeezed thermal state.
The condition for the quasiprobability distribution to be regular is 
used to test whether the states are either classical or nonclassical.

To compare with the observation, we calculated the power spectrum of the
gravitational wave perturbation with the nonclassical states from the
two-point correlation function. For the squeezed state, the power 
spectrum is corrected by $\left(\frac{H}{M}\right)^2$ which is too 
small to be detected from the observation. 
It is expected for linear order, $\frac{H}{M}$,
corrections to be detectable by the observations \cite{bergstrom02}.
And the squeeze parameter satisfies the condition for the states being
nonclassical. 
The power spectrum for the
squeezed coherent state consists of squeezed vacuum state contribution part
and classical part.  
The classical quantity $\mu_{kc}$, which is the expectation 
value of the gravitational 
field with respect to the squeezed coherent state, satisfies the classical
equation of motion (\ref{eom}).
The regular condition of 
quasiprobability distribution of the squeezed coherent state
does not depend on the amplitude of the squeezed coherent state, so
this state may be nonclassical as the squeezed vacuum state.
And for the 
squeezed thermal state, the power spectrum
 is modified by the thermal contribution
$(1+2\sinh^2 \theta(\beta_k))$. The nonclasscial condition depends on
the parameter $\beta$ and $\omega_k$.

Finally, the quasiprobability distribution measures the degree of the 
nonclassical behavior of states. This property will be used to investigate
the transition from quantum to classical perturbations during the inflation
for superhorizon scales.

\acknowledgements

We would like to thank D. N. Page and D. Pogosyan 
for useful discussions. We also
would like to appreciate the warm hospitality of the Theoretical
Physics Institute, University of Alberta.
This was supported by Korea Astronomy Observatory (KAO).

\appendix

\section{Quasiprobability Distribution} \label{quasiprobability}

We briefly review the properties of the quasiprobability distribution
\cite{cahill69}.
The ensemble average of the operator $F$ is represented by the density
operator
\begin{eqnarray}
\langle F \rangle = {\rm Tr}(\rho F).
\end{eqnarray}
This can be expressed in terms of the weight function $W(\alpha,s)$
by the integral form
\begin{eqnarray}
{\rm Tr} (\rho F)=\frac{1}{\pi}\int d^2 \alpha W(\alpha,s)f(\alpha,-s),
\end{eqnarray}
where $\alpha$ is a complex variable. Though $W(\alpha,s)$ can not exactly 
be interpreted as a probability distribution, it plays so closely the role of one that
it is referred to as a quasiprobability distribution. The $W(\alpha,s)$ is
identified with Glauber-Sudarshan ($P$), with the Wigner ($W$), and with
Husimi ($Q$) distribution corresponding  to the values $s=1, 0$, and $-1$,
respectively \cite{lutkenhaus95}.

Through complex Fourier transformations \cite{glauber69}, 
$W(\alpha,s)$ is expressed 
in terms of the characteristic function $\chi(\xi,s)$ as
\begin{eqnarray}
W(\alpha,s)=\frac{1}{\pi}\int e^{\alpha \xi^{\ast}-\alpha^{\ast}\xi}
\chi(\xi,s) d^2 \xi, 
\label{quasi_dist}
\end{eqnarray}
where
\begin{eqnarray}
\chi(\xi,s)={\rm Tr}(\rho \hat{D}(\xi))e^{\frac{s}{2}|\xi|^2},
\end{eqnarray}
with the displacement operator $\hat{D}(\xi)$.
The $W(\alpha,s)$ is normalized to unit through the trace of density operator
\begin{eqnarray}
1={\rm Tr}\rho = \frac{1}{\pi}\int W(\alpha,s) d^2 \alpha,
\end{eqnarray}
and satisfies the Hermiticity
\begin{eqnarray}
W(\alpha,s^{\ast})=W(\alpha,s)^{\ast},
\end{eqnarray}
which shows that the function $W(\alpha,s)$ is real for real values $s$.
Especially, the $P$ distribution is more useful in measuring the degree of
the nonclassical behavior. If the $P$ distribution of any state is more
singular than a delta function and is not positive definite,
this state is classified as a nonclassical state \cite{hillery85}. The $P$
distribution no more singular than a delta function will be said 
to regular.

The characteristic function of the squeezed coherent thermal states
is derived \cite{marian93}
\begin{eqnarray}
\chi(\xi,s)=\exp \left[-A |\xi|^2-\frac{1}{2}
(B^{\ast}\xi^2+B \xi^{\ast 2})+C^{\ast}\xi-C \xi^{\ast}\right], 
\label{character}
\end{eqnarray}
where 
\begin{eqnarray}
A &=& \frac{1-s}{2}+\bar{n}+(2\bar{n}+1)\sinh^2 r, \nonumber \\
B &=& -(2\bar{n}+1)e^{i\phi} \sinh r \cosh r, \nonumber \\
C &=& \lambda, \nonumber \\
\bar{n} &=& (e^{\beta \omega} -1)^{-1}.
\end{eqnarray}
The condition for the integration of the quasiprobability distribution
(\ref{quasi_dist}) to be regular 
with the characteristic function (\ref{character})
is 
\cite{marian93,kim89,lutkenhaus95}
\begin{eqnarray}
s < s_c \equiv (2\bar{n}+1) e^{-2r}. \label{condi}
\end{eqnarray}
Then the quasiprobability distribution is calculated for squeezed coherent
thermal states
\cite{marian93}
\begin{eqnarray}
W(\alpha,s) &=& (A^2 -|B|^2)^{-\frac{1}{2}} \nonumber \\
&\times&\exp 
\left[-\frac{1}{A^2-|B|^2}\left(A |\alpha-C|^2 +\frac{1}{2}[B^{\ast}
(\alpha-C)^2+B(\alpha^{\ast}-C^{\ast})]\right)\right]. \label{quasisol}
\end{eqnarray}
This result can be applied to the squeezed vacuum state by letting
$\lambda=0$ and $\bar{n}=0$, to the squeezed coherent state by $\bar{n}=0$,
and to the squeezed thermal state by $\lambda=0$. 
Depending on whether a state satisfies the condition (\ref{condi}),
it shows either a classical or nonclassical behavior.
This result simply extends to the two-mode squeezed state if the
characteristic function is given by the  exponential of a quadratic
form \cite{lutkenhaus95, soto83}.

\end{document}